\documentstyle[12pt,epsf]{article}
\newcommand{\be}{\begin{equation}}
\newcommand{\ee}{\end{equation}}

\newcommand{\beqa}{\begin{eqnarray}}
\newcommand{\eeqa}{\end{eqnarray}}
\newcommand{\nn}{\nonumber}

\newcommand{\eqref}[1]{(\ref{#1})}

\def\boxit#1{\vbox{\hrule\hbox{\vrule\kern8pt
\vbox{\hbox{\kern8pt}\hbox{\vbox{#1}}\hbox{\kern8pt}}
\kern8pt\vrule}\hrule}}
\def\mathboxit#1{\vbox{\hrule\hbox{\vrule\kern8pt\vbox{\kern8pt
\hbox{$\displaystyle #1$}\kern8pt}\kern8pt\vrule}\hrule}}

\def\IB{\relax\hbox{$\inbar\kern-.3em{\rm B}$}}
\def\IC{\relax\hbox{$\inbar\kern-.3em{\rm C}$}}
\def\ID{\relax\hbox{$\inbar\kern-.3em{\rm D}$}}
\def\IE{\relax\hbox{$\inbar\kern-.3em{\rm E}$}}
\def\IF{\relax\hbox{$\inbar\kern-.3em{\rm F}$}}
\def\IG{\relax\hbox{$\inbar\kern-.3em{\rm G}$}}
\def\IGa{\relax\hbox{${\rm I}\kern-.18em\Gamma$}}
\def\IH{\relax{\rm I\kern-.18em H}}
\def\IK{\relax{\rm I\kern-.18em K}}
\def\IL{\relax{\rm I\kern-.18em L}}
\def\IP{\relax{\rm I\kern-.18em P}}
\def\IR{\relax{\rm I\kern-.18em R}}
\def\IZ{\relax\ifmmode\mathchoice
{\hbox{\cmss Z\kern-.4em Z}}{\hbox{\cmss Z\kern-.4em Z}}
{\lower.9pt\hbox{\cmsss Z\kern-.4em Z}} {\lower1.2pt\hbox{\cmsss
Z\kern-.4em Z}}\else{\cmss Z\kern-.4em Z}\fi}

\def\II{\relax{\rm I\kern-.18em I}}

\def\CA {{\cal A}}

\def\CG {{\cal G}}

\def\CL {{\cal L}}

\begin{document}

\hfill  NRCPS-HE-05-56

\vspace{1cm}
%\begin{titlepage}
%\title{
\begin{center}
{\large ~\\ { Non-Abelian Tensor Gauge Fields  }\\
{\it and } \\
{ Higher-Spin Extension of Standard Model }\\

}%title ends

\vspace{1cm}
%\author{

{\sl George Savvidy\\
Demokritos National Research Center\\
Institute of Nuclear Physics\\
Ag. Paraskevi, GR-15310 Athens,Greece  \\
\centerline{\footnotesize\it E-mail: savvidy(AT)inp.demokritos.gr}
}%author ends
%}
%\date{}%in order NOT to write the date
%\maketitle
\end{center}
\vspace{60pt}

\centerline{{\bf Abstract}}

\vspace{12pt}

\noindent
  We suggest an extension of the Yang-Mills theory which includes non-Abelian tensor gauge fields.
The invariant Lagrangian is quadratic in
the field strength tensors and describes interaction of charged tensor gauge
bosons of arbitrary large integer spin $1,2,...$.
Non-Abelian tensor gauge fields can be viewed as a unique gauge field with values
in the infinite-dimensional current algebra associated with compact Lie group.
The full Lagrangian exhibits also enhanced local gauge invariance with
double number of gauge parameters which
allows to eliminate all negative norm states of the nonsymmetric second-rank tensor gauge field,
which describes therefore two polarizations of helicity-two massless charged
tensor gauge boson and the helicity-zero "axion". The geometrical interpretation
of the enhanced gauge symmetry with double number of gauge parameters is not yet
known. We suggest higher-spin extension of the electroweak theory and consider
creation processes of new tensor gauge bosons.

\newpage

\tableofcontents

%\end{abstract}
%\thispagestyle{empty}
%\end{titlepage}

\pagestyle{plain}
%\pagenumbering{roman}

\section{Introduction}

The non-Abelian local gauge invariance, which was formulated
by Yang and Mills in \cite{yang},
requires that all interactions must be invariant under
independent rotations of internal
charges at all space-time points.
The gauge principle allows very little arbitrariness: the interaction of matter
fields,  which carry non-commuting internal charges, and the nonlinear
self-interaction of gauge bosons are essentially fixed by the requirement
of local gauge invariance, very similar to the self-interaction of
gravitons in general relativity.

It is therefore appealing to extend the gauge principle, which was elevated by Yang and
Mills to a powerful constructive principle, so that it will define the
interaction of matter fields which carry
not only non-commutative internal charges, but
also arbitrary large spins. It seems that this will naturally
lead to a theory in which fundamental forces will be mediated by
integer-spin gauge quanta  and
that the Yang-Mills vector gauge boson will become a member of a bigger
family of tensor gauge bosons.

In the recent papers
\cite{Savvidy:2005fi,Savvidy:2005zm,Savvidy:dv,Savvidy:2003fx,Savvidy:2005fe}
 we extended the
gauge principle so that it enlarges the Yang-Mills group of local gauge transformations
and defines interaction of tensor gauge bosons of arbitrary large integer spins.
The extended non-Abelian gauge
transformations of the tensor gauge fields form {\it a new large group
which has a natural geometrical interpretation in terms of extended current
algebra associated with compact Lie group G}.
On this large group one can define field strength tensors,
which are transforming homogeneously with respect to the extended
gauge transformations.  The invariant Lagrangian is quadratic in
the field strength tensors and describes interaction of tensor gauge
fields of arbitrary large integer spin $1,2,...$.
It was also demonstrated that the total Lagrangian
exhibits enhanced local gauge invariance with double number of gauge parameters.
This allows to eliminate all negative norm states of the nonsymmetric second rank
tensor gauge field $A_{\mu\lambda}$, which describes therefore
two polarizations of helicity-two massless charged tensor gauge boson and
the helicity-zero "axion". We suggest higher-spin extension of the
electroweak theory and consider creation processes of new tensor gauge bosons.

The early investigation of higher-spin representations of the Poincar\'e
algebra and of the corresponding field equations is due to Majorana,
Dirac and Wigner. The theory of massive particles
of higher spin was further developed by Fierz and Pauli \cite{fierzpauli} and
Rarita and Schwinger \cite{rarita}. The Lagrangian and
S-matrix formulations of {\it free field
theory} of massive and massless fields with higher spin
have been completely constructed in
\cite{yukawa1,schwinger,Weinberg:1964cn,
chang,singh,singh1,fronsdal,fronsdal1}.
The problem of {\it introducing interaction} appears to be much more complex
\cite{fronsdal2} and met enormous difficulties for spin fields higher than two
\cite{berends}. The first positive result in this direction was
the light-front construction of the cubic
interaction term for the massless field of helicity $\pm \lambda$ in
\cite{Bengtsson:1983pd,Bengtsson:1983pg}.

\section{Extended Gauge Transformations}

In our approach the gauge fields are defined as rank-$(s+1)$ tensors
\cite{Savvidy:2005fi,Savvidy:2005zm,Savvidy:dv,Savvidy:2003fx,Savvidy:2005fe}
$$
A^{a}_{\mu\lambda_1 ... \lambda_{s}}(x),~~~~~s=0,1,2,...
$$
and are totally symmetric with respect to the
indices $  \lambda_1 ... \lambda_{s}  $. {\it A priory} the tensor fields
have no symmetries with
respect to the first index  $\mu$. The index $a$ numerates the generators $L^a$
of the Lie algebra $\breve{g}$ of a {\it compact}\footnote{The algebra $\breve{g}$
possesses an orthogonal
basis in which the structure constant $f^{abc}$ are totally antisymmetric.}
Lie group G.
One can think of these tensor fields as appearing in the
expansion of the extended gauge field $\CA_{\mu}(x,e)$ over the tangent space-like unit vector
$e_{\lambda}$ \cite{Savvidy:2005fi,Savvidy:2005zm,Savvidy:dv,Savvidy:2003fx,Savvidy:2005fe}
\be\label{gaugefield}
\CA_{\mu}(x,e) = \sum^{\infty}_{s=0}~{1\over s!} ~
A^{a}_{\mu\lambda_1 ... \lambda_{s}}(x) ~L^a e_{\lambda_1}...e_{\lambda_s}.
\ee
The gauge field $A^{a}_{\mu\lambda_1 ... \lambda_{s}}$ carry
indices $a,\lambda_1, ..., \lambda_{s}$ labeling the generators of {\it extended current
algebra $\CG$ associated with compact Lie group G.} It has infinite many generators
$L^{a}_{\lambda_1 ... \lambda_{s}} = L^a e_{\lambda_1}...e_{\lambda_s}$ and
the corresponding algebra is given by the commutator
\be
[L^{a}_{\lambda_1 ... \lambda_{s}}, L^{b}_{\rho_1 ... \rho_{k}}]=if^{abc}
L^{c}_{\lambda_1 ... \lambda_{s}\rho_1 ... \rho_{k}}.
\ee

{\it The extended non-Abelian gauge transformations of the tensor gauge fields are defined
by the following equations } \cite{Savvidy:2005zm}:
\beqa\label{polygauge}
\delta A^{a}_{\mu} &=& ( \delta^{ab}\partial_{\mu}
+g f^{acb}A^{c}_{\mu})\xi^b ,~~~~~\\
\delta A^{a}_{\mu\nu} &=&  ( \delta^{ab}\partial_{\mu}
+  g f^{acb}A^{c}_{\mu})\xi^{b}_{\nu} + g f^{acb}A^{c}_{\mu\nu}\xi^{b},\nonumber\\
\delta A^{a}_{\mu\nu \lambda}& =&  ( \delta^{ab}\partial_{\mu}
+g f^{acb} A^{c}_{\mu})\xi^{b}_{\nu\lambda} +
g f^{acb}(  A^{c}_{\mu  \nu}\xi^{b}_{\lambda } +
A^{c}_{\mu \lambda }\xi^{b}_{ \nu}+
A^{c}_{\mu\nu\lambda}\xi^{b}),\nn\\
.........&.&............................\nn
\eeqa

These extended gauge transformations
generate a closed algebraic structure. To see that, one should compute the
commutator of two extended gauge transformations $\delta_{\eta}$ and $\delta_{\xi}$
of parameters $\eta$ and $\xi$.
The commutator of two transformations can be expressed in the form \cite{Savvidy:2005zm}
\be\label{gaugecommutator}
[~\delta_{\eta},\delta_{\xi}]~A_{\mu\lambda_1\lambda_2 ...\lambda_s} ~=~
-i g~ \delta_{\zeta} A_{\mu\lambda_1\lambda_2 ...\lambda_s}
\ee
and is again an extended gauge transformation with the gauge parameters
$\{\zeta\}$ which are given by the matrix commutators
\beqa\label{gaugealgebra}
\zeta&=&[\eta,\xi]\\
\zeta_{\lambda_1}&=&[\eta,\xi_{\lambda_1}] +[\eta_{\lambda_1},\xi]\nn\\
\zeta_{\nu\lambda} &=& [\eta,\xi_{\nu\lambda}] +  [\eta_{\nu},\xi_{\lambda}]
+ [\eta_{\lambda},\xi_{\nu}]+[\eta_{\nu\lambda},\xi],\nn\\
......&.&..........................\nn
\eeqa
{\it The generalized field strengths  are defined as} \cite{Savvidy:2005zm}
\beqa\label{fieldstrengthparticular}
G^{a}_{\mu\nu} &=&
\partial_{\mu} A^{a}_{\nu} - \partial_{\nu} A^{a}_{\mu} +
g f^{abc}~A^{b}_{\mu}~A^{c}_{\nu},\\
G^{a}_{\mu\nu,\lambda} &=&
\partial_{\mu} A^{a}_{\nu\lambda} - \partial_{\nu} A^{a}_{\mu\lambda} +
g f^{abc}(~A^{b}_{\mu}~A^{c}_{\nu\lambda} + A^{b}_{\mu\lambda}~A^{c}_{\nu} ~),\nn\\
G^{a}_{\mu\nu,\lambda\rho} &=&
\partial_{\mu} A^{a}_{\nu\lambda\rho} - \partial_{\nu} A^{a}_{\mu\lambda\rho} +
g f^{abc}(~A^{b}_{\mu}~A^{c}_{\nu\lambda\rho} +
 A^{b}_{\mu\lambda}~A^{c}_{\nu\rho}+A^{b}_{\mu\rho}~A^{c}_{\nu\lambda}
 + A^{b}_{\mu\lambda\rho}~A^{c}_{\nu} ~),\nn\\
 ......&.&............................................\nn
\eeqa
and transform homogeneously with respect to the extended
gauge transformations (\ref{polygauge}). The field strength tensors are
antisymmetric in their first two indices and are totaly symmetric with respect to the
rest of the indices.

The inhomogeneous extended gauge transformation (\ref{polygauge})
induces the homogeneous gauge
transformation of the corresponding field strength
(\ref{fieldstrengthparticular}) of the form \cite{Savvidy:2005zm}
\beqa\label{fieldstrenghparticulartransformation}
\delta G^{a}_{\mu\nu}&=& g f^{abc} G^{b}_{\mu\nu} \xi^c \\
\delta G^{a}_{\mu\nu,\lambda} &=& g f^{abc} (~G^{b}_{\mu\nu,\lambda} \xi^c
+ G^{b}_{\mu\nu} \xi^{c}_{\lambda}~),\nonumber\\
\delta G^{a}_{\mu\nu,\lambda\rho} &=& g f^{abc}
(~G^{b}_{\mu\nu,\lambda\rho} \xi^c
+ G^{b}_{\mu\nu,\lambda} \xi^{c}_{\rho} +
G^{b}_{\mu\nu,\rho} \xi^{c}_{\lambda} +
G^{b}_{\mu\nu} \xi^{c}_{\lambda\rho}~)\nn\\
......&.&..........................,\nn
\eeqa
The field strength tensors are
antisymmetric in their first two indices and are totaly symmetric with respect to the
rest of the indices.
The symmetry properties of the field strength  $G^{a}_{\mu\nu,\lambda_1 ... \lambda_s}$
remain invariant in the course of this transformation.
By induction the entire construction
can be generalized to include tensor fields of any rank s \cite{Savvidy:2005fi,Savvidy:2005zm}.

\section{The First Gauge Invariant Lagrangian}

The gauge invariant Lagrangian  now can be formulated in the form \cite{Savvidy:2005zm}
\beqa\label{fulllagrangian1}
{{\cal L}}_{s+1}&=&-{1\over 4} ~
G^{a}_{\mu\nu, \lambda_1 ... \lambda_s}~
G^{a}_{\mu\nu, \lambda_{1}...\lambda_{s}} +.......\nonumber\\
&=& -{1\over 4}\sum^{2s}_{i=0}~a^{s}_i ~
G^{a}_{\mu\nu, \lambda_1 ... \lambda_i}~
G^{a}_{\mu\nu, \lambda_{i+1}...\lambda_{2s}}
(\sum_{P} \eta^{\lambda_{i_1} \lambda_{i_2}} .......
\eta^{\lambda_{i_{2s-1}} \lambda_{i_{2s}}})~,
\eeqa
where the sum $\sum_P$ runs over all nonequal permutations of $i's$, in total $(2s-1)!!$
terms. For the low
values of $s=0,1,2,...$ the numerical coefficients $$a^{s}_i = {s!\over i!(2s-i)!}$$
are: $a^{0}_0=1;~~a^{1}_1 =1,a^{1}_0 =a^{1}_2 =1/2;~~
a^{2}_2 =1/2,a^{2}_1 =a^{2}_3 =1/3,a^{2}_0 =a^{2}_4 =1/12;$ and so on.
In order to describe fixed rank-$(s+1)$ gauge field
one should have  at disposal all gauge fields
up to the rank $2s+1$.
In order to make all tensor gauge fields dynamical one should add
the corresponding kinetic terms. Thus the invariant
Lagrangian describing dynamical tensor gauge bosons of all ranks
has the form
\be\label{fulllagrangian2}
{{\cal L}} = \sum^{\infty}_{s=1}~ g_s {{\cal L}}_s~.
\ee
The first three terms of the invariant Lagrangian have
the following form \cite{Savvidy:2005zm}:
\beqa\label{firstthreeterms}
{{\cal L}} =  {{\cal L}}_1 +  {{\cal L}}_2 +{{\cal L}}_3 +... =-{1\over 4}G^{a}_{\mu\nu}
G^{a}_{\mu\nu}
-{1\over 4}G^{a}_{\mu\nu,\lambda}G^{a}_{\mu\nu,\lambda}
-{1\over 4}G^{a}_{\mu\nu}G^{a}_{\mu\nu,\lambda\lambda}~-~~~~~~~~~~\nn\\
-{1\over 4}G^{a}_{\mu\nu,\lambda\rho}G^{a}_{\mu\nu,\lambda\rho}
-{1\over 8}G^{a}_{\mu\nu ,\lambda\lambda}G^{a}_{\mu\nu ,\rho\rho}
-{1\over 2}G^{a}_{\mu\nu,\lambda}  G^{a}_{\mu\nu ,\lambda \rho\rho}
-{1\over 8}G^{a}_{\mu\nu}  G^{a}_{\mu\nu ,\lambda \lambda\rho\rho}~+..,
\eeqa
where the first term is the Yang-Mills Lagrangian and the
second and the third ones describe the tensor gauge fields $A^{a}_{\mu\nu},
A^{a}_{\mu\nu\lambda}$ and so on.
It is important that:  i) {\it the Lagrangian does not
contain higher derivatives of tensor gauge fields
ii) all interactions take place
through the three- and four-particle exchanges with dimensionless
coupling constant  iii) the complete Lagrangian contains all higher-rank
tensor gauge fields and should not be truncated}.

\section{Geometrical Interpretation}

Let us consider a possible geometrical interpretation of the
above construction. Introducing tangent space-like unit vector $e_{\mu}$ and considering it
as a second variable we can introduce the extended gauge parameter
$\xi^{a}_{\lambda_1 ... \lambda_{s}}(x)$ and the generators
$L^{a}_{\lambda_1...\lambda_s} = L^a e_{\lambda_1}...e_{\lambda_s}$
\cite{Savvidy:2005fi,Savvidy:2005zm,Savvidy:dv,Savvidy:2003fx,Savvidy:2005fe}
\be
\xi(x,e)=   \sum^{\infty}_{s=0}~{1\over s!} ~
\xi^{a}_{\lambda_1 ... \lambda_{s}}(x) ~L^a   e_{\lambda_1}...e_{\lambda_s}
\ee
and define the gauge transformation of the extended gauge field
$\CA_{\mu}(x,e)$   as in (\ref{polygauge})
\be\label{extendedgaugetransformation}
\CA^{'}_{\mu}(x,e) = U(\xi)  \CA_{\mu}(x,e) U^{-1}(\xi) -{i\over g}
\partial_{\mu}U(\xi) ~U^{-1}(\xi),
\ee
where the unitary transformation matrix
is given by the expression
$
U(\xi) = exp\{ig \xi(x,e) \}.
$
This allows to construct the extended field strength tensor of the form
(\ref{fieldstrengthparticular})
\be\label{fieldstrengthgeneral}
\CG_{\mu\nu}(x,e) = \partial_{\mu} \CA_{\nu}(x,e) - \partial_{\nu} \CA_{\mu}(x,e) -
i g [ \CA_{\mu}(x,e)~\CA_{\nu}(x,e)]
\ee
using the commutator of the covariant derivatives
$
\nabla^{ab}_{\mu} = (\partial_{\mu}-ig \CA_{\mu}(x,e))^{ab}
$
of a standard form
$
[\nabla_{\mu}, \nabla_{\nu}]^{ab} = g f^{acb} \CG^{c}_{\mu\nu}~,
$
so that
\be\label{fieldstrenghthtransformation}
\CG^{'}_{\mu\nu}(x,e)) = U(\xi)  \CG_{\mu\nu}(x,e) U^{-1}(\xi).
\ee
The  invariant Lagrangian density is given by the expression
\be\label{lagrangdensity}
{{\cal L}}(x,e) =  \CG^{a}_{\mu\nu}(x,e)\CG^{a}_{\mu\nu}(x,e),
\ee
as one can be convinced computing its variation with respect to the
extended gauge transformation (\ref{polygauge}),(\ref{extendedgaugetransformation})
and (\ref{fieldstrenghparticulartransformation}),(\ref{fieldstrenghthtransformation})
$$
\delta {{\cal L}}(x,e) = 2 \CG^{a}_{\mu\nu}(x,e)~ g f^{acb}~
\CG^{c}_{\mu\nu}(x,e) ~\xi^{b}(x,e) =0.
$$
The Lagrangian density (\ref{lagrangdensity}) allows to extract
{\it gauge invariant, totally symmetric, tensor densities $\CL_{\lambda_1 ... \lambda_{s}}(x)$}
using expansion with respect to the
vector variable $e$
\be
\CL(x,e) = \sum^{\infty}_{s=0}~{1\over s!} ~
\CL_{\lambda_1 ... \lambda_{s}}(x) ~ e_{\lambda_1}...e_{\lambda_s} .
\ee
In particular the expansion term which is quadratic in powers of $e$ is
(see the next section for explicit variation (\ref{explicitvariation}))
\be
(\CL_2 )_{\lambda\rho} = -{1\over 4}G^{a}_{\mu\nu,\lambda}G^{a}_{\mu\nu,\rho}
-{1\over 4}G^{a}_{\mu\nu}G^{a}_{\mu\nu,\lambda\rho}
\ee
and defines a unique Lorentz invariant Lagrangian which can be constructed from the above
tensor, that is the Lagrangian $\CL_2$
$$
{{\cal L}}_2 =-{1\over 4}G^{a}_{\mu\nu,\lambda}G^{a}_{\mu\nu,\lambda}
-{1\over 4}G^{a}_{\mu\nu}G^{a}_{\mu\nu,\lambda\lambda}.
$$

The whole construction can be viewed as an extended vector bundle X
on which the gauge field $\CA^{a}_{\mu}(x,e)$ is a connection. The
gauge field $A^{a}_{\mu\lambda_1 ... \lambda_{s}}$ carry
indices $a,\lambda_1, ..., \lambda_{s}$ which label the generators of the {\it extended current
algebra $\CG$ associated with the compact Lie group G}. It has infinite many generators
$L^{a}_{\lambda_1 ... \lambda_{s}} = L^a e_{\lambda_1}...e_{\lambda_s}$ with commutator
\be
[L^{a}_{\lambda_1 ... \lambda_{s}}, L^{b}_{\rho_1 ... \rho_{k}}]=if^{abc}
L^{c}_{\lambda_1 ... \lambda_{s}\rho_1 ... \rho_{k}}.
\ee
Thus we have vector bundle whose structure group
is an extended gauge group $\CG$ with group elements $U(\xi)=exp(i \xi(e) )$,
where $\xi(e)=  \sum_s ~
\xi^{a}_{\lambda_1 ... \lambda_{s}} ~L^a  e_{\lambda_1}...e_{\lambda_s}$
and the composition law (\ref{gaugealgebra}).
In contrast, in Kac-Moody current algebra
the generators depend on the complex variable $L^{a}_n = L^a z^n$ (see also \cite{minnes})
$$
[L^{a}_n ,L^{b}_m] = if^{abc} L^{c}_{n+m}.
$$
In the next section we shall see that, there exist a second invariant Lagrangian
${{\cal L}}^{'}$ which can be constructed in terms of extended
field strength tensors (\ref{fieldstrengthparticular}) and the total Lagrangian is a
linear sum of the two Lagrangians $ c ~{{\cal L}} + c^{'}~ {{\cal L}}^{'} $.

\section{The Second Gauge Invariant Lagrangian }

Indeed the Lagrangian (\ref{fulllagrangian1}), (\ref{fulllagrangian2})
and (\ref{firstthreeterms}) is not the
most general Lagrangian which can be constructed in terms
of the above field strength tensors
(\ref{fieldstrengthparticular}) and (\ref{fieldstrengthgeneral}).
As we shall see there exists a second invariant Lagrangian ${{\cal L}}^{'}$ (\ref{secondseries}),
(\ref{secondseriesdensities}) and (\ref{secondfulllagrangian})
which can be constructed in terms of extended
field strength tensors (\ref{fieldstrengthparticular}) and the total Lagrangian is a
linear sum of the two Lagrangians $ c ~{{\cal L}} + c^{'}~ {{\cal L}}^{'} $.
In particular for the second-rank
tensor gauge field $A^{a}_{\mu\lambda}$ the total Lagrangian is a
sum of two Lagrangians $c{{\cal L}}_{2}+c^{'}{{\cal L}}^{'}_{2} $ and,
with specially chosen coefficients $\{c,c^{'}\}$, it exhibits an enhanced
gauge invariance (\ref{largegaugetransformation}),(\ref{freedoublepolygaugesymmetric})
with double number of gauge parameters, which allows to eliminate negative norm
polarizations of the nonsymmetric second-rank tensor gauge field $A_{\mu\lambda}$.
The geometrical interpretation
of the enhanced gauge symmetry with double number of gauge parameters is not yet
known.

Let us consider the gauge invariant tensor density of the form
\cite{Savvidy:2005fi,Savvidy:2005zm,Savvidy:2003fx}
\be\label{secondseries}
{{\cal L}}^{'}_{\rho_1\rho_2}(x,e) =  {1\over 4}
\CG^{a}_{\mu\rho_1}(x,e)\CG^{a}_{\mu\rho_2}(x,e).
\ee
It is gauge invariant because its variation is also equal to zero
$$
\delta {{\cal L}}^{'}_{\rho_1\rho_2}(x,e) ={1\over 4}g f^{acb}~
\CG^{c}_{\mu\rho_1}(x,e) ~\xi^{b}(x,e)\CG^{a}_{\mu\rho_2}(x,e)+
{1\over 4}\CG^{a}_{\mu\rho_1}(x,e)~ g f^{acb}~
\CG^{c}_{\mu\rho_2}(x,e) ~\xi^{b}(x,e) =0.
$$
The Lagrangian density (\ref{secondseries}) generate the
second series of {\it gauge invariant tensor densities
$(\CL^{'}_{\rho_1\rho_2})_{\lambda_1 ... \lambda_{s}}(x)$}
when we expand it in powers of the vector variable $e$
\be\label{secondseriesdensities}
{{\cal L}}^{'}_{\rho_1\rho_2}(x,e) = \sum^{\infty}_{s=0}~{1\over s!} ~
(\CL^{'}_{\rho_1\rho_2})_{\lambda_1 ... \lambda_{s}}(x) ~ e_{\lambda_1}...e_{\lambda_s} .
\ee
Using contraction of these tensor densities the gauge invariant Lagrangians can
be formulated in the form
\beqa\label{secondfulllagrangian}
{{\cal L}}^{'}_{s+1}&=&{1\over 4} ~
G^{a}_{\mu\nu,\rho\lambda_3  ... \lambda_{s+1}}~
G^{a}_{\mu\rho,\nu\lambda_{3} ...\lambda_{s+1}} +{1\over 4} ~
G^{a}_{\mu\nu,\nu\lambda_3  ... \lambda_{s+1}}~
G^{a}_{\mu\rho,\rho\lambda_{3} ...\lambda_{s+1}} +.......\nonumber\\
&=& {1\over 4}\sum^{2s+1}_{i=1}~{ a^{s}_{i-1}\over s}  ~
G^{a}_{\mu\lambda_1,\lambda_2  ... \lambda_i}~
G^{a}_{\mu\lambda_{i+1},\lambda_{i+2} ...\lambda_{2s+2}}
(\sum^{'}_{P} \eta^{\lambda_{i_1} \lambda_{i_2}} .......
\eta^{\lambda_{i_{2s+1}} \lambda_{i_{2s+2}}})~,
\eeqa
where the sum $\sum^{'}_P$ runs over all nonequal permutations of $i's$, with exclusion
of the terms which contain $\eta^{\lambda_{1},\lambda_{i+1}}$.

It is also instructive  to construct these Lagrangian densities explicitly.
The invariance of the first Lagrangian ${{\cal L}}_2$
$$
{{\cal L}}_2 =-{1\over 4}G^{a}_{\mu\nu,\lambda}G^{a}_{\mu\nu,\lambda}
-{1\over 4}G^{a}_{\mu\nu}G^{a}_{\mu\nu,\lambda\lambda}
$$
in (\ref{fulllagrangian1}), (\ref{fulllagrangian2}) and (\ref{firstthreeterms})
was demonstrated in \cite{Savvidy:2005zm} by
calculating its variation with respect to the gauge
transformation (\ref{polygauge}) and (\ref{fieldstrenghparticulartransformation}).
Indeed, its explicit variation is equal to zero
\beqa\label{explicitvariation}
\delta \CL_2   =
&-&{1\over 4} G^{a}_{\mu\nu,\lambda} g f^{abc} (G^{b}_{\mu\nu,\lambda} \xi^c  +
G^{b}_{\mu\nu} \xi^{c}_{\lambda})
-{1\over 4}  g f^{abc} (G^{a}_{\mu\nu,\lambda}\xi^c +
G^{b}_{\mu\nu} \xi^{c}_{\lambda}   )G^{b}_{\mu\nu,\lambda}   \nn\\
&-&{1\over 4} g f^{abc} G^{b}_{\mu\nu} \xi^c G^{a}_{\mu\nu,\lambda\lambda}\nonumber\\
&-&{1\over 4} G^{a}_{\mu\nu} g f^{abc} (G^{b}_{\mu\nu,\lambda\lambda} \xi^c  +
G^{b}_{\mu\nu, \lambda} \xi^{c}_{\lambda}+G^{b}_{\mu\nu, \lambda} \xi^{c}_{\lambda}+
G^{b}_{\mu\nu} \xi^{c}_{\lambda \lambda})=0 .
\eeqa
As we have seen above a general consideration shows that the Lagrangian ${{\cal L}}_{2}$
is not a unique one and that there exist a second invariant Lagrangian ${{\cal L}}^{'}_2$.
Let us construct this Lagrangian density explicitly. First notice that there
exist additional Lorentz invariant
quadratic forms which can be constructed by the corresponding field
strength tensors. They are \cite{Savvidy:2005fi}
$$
G^{a}_{\mu\nu,\lambda}G^{a}_{\mu\lambda,\nu},~~~
G^{a}_{\mu\nu,\nu}G^{a}_{\mu\lambda,\lambda},~~~
G^{a}_{\mu\nu}G^{a}_{\mu\lambda,\nu\lambda}.
$$
Calculating  the variation of each of these terms with respect to
the gauge transformation (\ref{polygauge}) and (\ref{fieldstrenghparticulartransformation})
one can get convinced that a particular linear combination
\be\label{actiontwoprime}
{{\cal L}}^{'}_2 =  {1\over 4}G^{a}_{\mu\nu,\lambda}G^{a}_{\mu\lambda,\nu}
+{1\over 4}G^{a}_{\mu\nu,\nu}G^{a}_{\mu\lambda,\lambda}
+{1\over 2} G^{a}_{\mu\nu}G^{a}_{\mu\lambda,\nu\lambda}
\ee
forms an invariant Lagrangian and coincides with (\ref{secondfulllagrangian}) for
s=1. Indeed the variation of the Lagrangian ${{\cal L}}^{'}_2$
under the gauge transformation (\ref{fieldstrenghparticulartransformation}) is equal to zero:
\beqa
\delta {{\cal L}}^{'}_2 =
&+&{1\over 4} G^{a}_{\mu\nu,\lambda} g f^{abc} (G^{b}_{\mu\lambda,\nu} \xi^c  +
G^{b}_{\mu\lambda} \xi^{c}_{\nu})
+{1\over 4} g f^{abc} (G^{b}_{\mu\nu,\lambda} \xi^c  +
G^{b}_{\mu\nu} \xi^{c}_{\lambda}) G^{a}_{\mu\lambda,\nu}    \nn\\
&+& {1\over 2}G^{a}_{\mu\nu,\nu}g f^{abc} (G^{b}_{\mu\lambda,\lambda}  \xi^c  +
G^{b}_{\mu\lambda} \xi^{c}_{\lambda})\nn\\
&+& {1\over 2}g f^{abc} G^{b}_{\mu\nu} \xi^c G^{a}_{\mu\lambda,\nu\lambda}\nonumber\\
&+& {1\over 2} G^{a}_{\mu\nu} g f^{abc} (G^{b}_{\mu\lambda,\nu\lambda} \xi^c  +
G^{b}_{\mu\lambda,\nu } \xi^{c}_{\lambda}+
G^{b}_{\mu\lambda,\lambda } \xi^{c}_{\nu}+
G^{b}_{\mu\lambda} \xi^{c}_{\nu \lambda})=0 .\nonumber
\eeqa
As a result we have two invariant Lagrangians
${{\cal L}}_2$ and ${{\cal L}}^{'}_2$ and the general Lagrangian is a
linear combination of these two Lagrangians
$
{{\cal L}}_2 + c {{\cal L}}^{'}_2 ,
$
where c is an arbitrary constant.

{\it Our aim now is to demonstrate  that
if $c=1$ then we shall have enhanced local gauge invariance
(\ref{largegaugetransformation}),(\ref{freedoublepolygaugesymmetric}) of the Lagrangian
${{\cal L}}_2 + {{\cal L}}^{'}_2$ with double number of gauge parameters.
This allows to eliminate all negative norm states of the nonsymmetric second-rank
tensor gauge field $A^{a}_{\mu \lambda}$, which describes therefore two polarizations
of helicity-two massless charged tensor gauge boson and
of the helicity-zero "axion".}

\section{Enhancement of Extended Gauge Transformations }

Indeed, let us consider the
situation at the linearized level when the gauge coupling constant g is equal to zero.
The free part of the ${{\cal L}}_2$ Lagrangian is
$$
{{\cal L}}^{free}_2 ={1 \over 2} A^{a}_{\alpha\acute{\alpha}}
(\eta_{\alpha\gamma}\eta_{\acute{\alpha}\acute{\gamma}}\partial^{2} -
\eta_{\acute{\alpha}\acute{\gamma}} \partial_{\alpha} \partial_{\gamma} )
A^{a}_{\gamma\acute{\gamma}} =
{1 \over 2} A^{a}_{\alpha\acute{\alpha}}
H_{\alpha\acute{\alpha}\gamma\acute{\gamma}} A^{a}_{\gamma\acute{\gamma}} ,
$$
where the quadratic form in the momentum representation has the form
$$
H_{\alpha\acute{\alpha}\gamma\acute{\gamma}}(k)=
-(k^2 \eta_{\alpha\gamma} -k_{\alpha}k_{\gamma})
\eta_{\acute{\alpha}\acute{\gamma}} =
 H_{\alpha \gamma }(k) \eta_{\acute{\alpha}\acute{\gamma}},
$$
is obviously invariant with respect to the gauge
transformation $\delta A^{a}_{\mu\lambda} =\partial_{\mu} \xi^{a}_{\lambda}$,
but it is not invariant with respect to the alternative gauge transformations
$\delta A^{a}_{\mu \lambda} =\partial_{\lambda} \eta^{a}_{\mu}$. This can be
seen, for example, from the following relations in the momentum representation:
\be\label{currentdivergence}
k_{\alpha}H_{\alpha\acute{\alpha}\gamma\acute{\gamma}}(k)=0,~~~
k_{\acute{\alpha}}H_{\alpha\acute{\alpha}\gamma\acute{\gamma}}(k)=
-(k^2 \eta_{\alpha\gamma} - k_{\alpha}k_{\gamma})k_{\acute{\gamma}} \neq 0 .
\ee
Let us consider now the free part of the second Lagrangian
\beqa
{{\cal L}}^{' free}_{2} ={1 \over 4} A^{a}_{\alpha\acute{\alpha}}
(-\eta_{\alpha\acute{\gamma}}\eta_{\acute{\alpha}\gamma}\partial^{2} -
\eta_{\alpha\acute{\alpha}}\eta_{\gamma\acute{\gamma}}\partial^{2}
+\eta_{\alpha\acute{\gamma}} \partial_{\acute{\alpha}} \partial_{\gamma}
+\eta_{\acute{\alpha}\gamma} \partial_{\alpha} \partial_{\acute{\gamma}}
+\eta_{\alpha\acute{\alpha}} \partial_{\gamma} \partial_{\acute{\gamma}}+\nn\\
+\eta_{\gamma\acute{\gamma}} \partial_{\alpha} \partial_{\acute{\alpha}}
-2\eta_{\alpha\gamma} \partial_{\acute{\alpha}} \partial_{\acute{\gamma}})
A^{a}_{\gamma\acute{\gamma}}=
{1 \over 2} A^{a}_{\alpha\acute{\alpha}}
H^{~'}_{\alpha\acute{\alpha}\gamma\acute{\gamma}} A^{a}_{\gamma\acute{\gamma}},
\eeqa
where
$$
H^{'}_{\alpha\acute{\alpha}\gamma\acute{\gamma}}(k)=
{1 \over 2}(\eta_{\alpha\acute{\gamma}}\eta_{\acute{\alpha}\gamma}
+\eta_{\alpha\acute{\alpha}}\eta_{\gamma\acute{\gamma}})k^2
-{1 \over 2}(\eta_{\alpha\acute{\gamma}}k_{\acute\alpha}k_{\gamma}
+\eta_{\acute\alpha\gamma}k_{\alpha}k_{\acute{\gamma}}
+\eta_{\alpha\acute\alpha}k_{\gamma}k_{\acute{\gamma}}
+\eta_{\gamma\acute{\gamma}}k_{\alpha}k_{\acute\alpha}
-2\eta_{\alpha\gamma}k_{\acute\alpha}k_{\acute{\gamma}}).
$$
It is again invariant with respect to the gauge
transformation $\delta A^{a}_{\mu\lambda} =\partial_{\mu} \xi^{a}_{\lambda}$,
but it is not invariant with respect to the gauge transformations
$\delta A^{a}_{\mu \lambda} =\partial_{\lambda} \eta^{a}_{\mu}$ as one can
see from analogous relations
\be\label{currentdivergenceprime}
k_{\alpha}H^{'}_{\alpha\acute{\alpha}\gamma\acute{\gamma}}(k)=0,~~~
k_{\acute{\alpha}}H^{'}_{\alpha\acute{\alpha}\gamma\acute{\gamma}}(k)=
(k^2 \eta_{\alpha\gamma} -k_{\alpha}k_{\gamma})k_{\acute{\gamma}} \neq 0 .
\ee
As it is obvious from (\ref{currentdivergence}) and
(\ref{currentdivergenceprime}), the total Lagrangian
${{\cal L}}^{free}_2 + {{\cal L}}^{' free}_2$ now  poses new enhanced
invariance with respect to the larger, eight parameter, gauge transformations
\be\label{largegaugetransformation}
\delta A^{a}_{\mu \lambda} =\partial_{\mu} \xi^{a}_{\lambda}+
\partial_{\lambda} \eta^{a}_{\mu} +...,
\ee
where $\xi^{a}_{\lambda}$ and $\eta^{a}_{\mu}$ are eight arbitrary functions, because
\be\label{zeroderivatives}
k_{\alpha}(H_{\alpha\acute{\alpha}\gamma\acute{\gamma}}+
H^{'}_{\alpha\acute{\alpha}\gamma\acute{\gamma}})=0,~~~
k_{\acute{\alpha}}(H_{\alpha\acute{\alpha}\gamma\acute{\gamma}}+
H^{'}_{\alpha\acute{\alpha}\gamma\acute{\gamma}})=0 .
\ee
Thus our free part of the Lagrangian is
\beqa\label{totalfreelagrangian}
{{\cal L}}^{tot~free}_{2} =&-&{1 \over 2}\partial_{\mu}
A^{a}_{\nu \lambda}\partial_{\mu} A^{a}_{\nu \lambda}
+{1 \over 2}\partial_{\mu} A^{a}_{\nu \lambda}\partial_{\nu} A^{a}_{\mu \lambda}+
\nn\\
&+&{1 \over 4} \partial_{\mu} A^{a}_{\nu \lambda} \partial_{\mu } A^{a}_{\lambda\nu}
-{1 \over 4} \partial_{\mu} A^{a}_{\nu \lambda} \partial_{\lambda} A^{a}_{\mu \nu}
-{1 \over 4}\partial_{\nu} A^{a}_{\mu \lambda} \partial_{\mu} A^{a}_{\lambda\nu }
+{1 \over 4} \partial_{\nu } A^{a}_{\mu\lambda} \partial_{\lambda} A^{a}_{\mu \nu}
\nn\\
&+&{1 \over 4}\partial_{\mu} A^{a}_{\nu \nu}\partial_{\mu} A^{a}_{\lambda\lambda}
-{1 \over 2}\partial_{\mu} A^{a}_{\nu \nu} \partial_{\lambda} A^{a}_{\mu\lambda}
+{1 \over 4}\partial_{\nu } A^{a}_{\mu\nu}\partial_{\lambda} A^{a}_{\mu\lambda}
\eeqa
or, in equivalent form, it is
\beqa\label{totfreelagrangianalternativeform}
{{\cal L}}^{tot~free}_{2} ={1 \over 2} A^{a}_{\alpha\acute{\alpha}}
\{(\eta_{\alpha\gamma}\eta_{\acute{\alpha}\acute{\gamma}}
-{1\over 2}\eta_{\alpha\acute{\gamma}}\eta_{\acute{\alpha}\gamma}
-{1\over 2}\eta_{\alpha\acute{\alpha}}\eta_{\gamma\acute{\gamma}})
\partial^{2}
-\eta_{\acute{\alpha}\acute{\gamma}} \partial_{\alpha} \partial_{\gamma}
-\eta_{\alpha\gamma} \partial_{\acute{\alpha}} \partial_{\acute{\gamma}}+\nn\\
+{1\over 2}(\eta_{\alpha\acute{\gamma}} \partial_{\acute{\alpha}} \partial_{\gamma}
+\eta_{\acute{\alpha}\gamma} \partial_{\alpha} \partial_{\acute{\gamma}}
+\eta_{\alpha\acute{\alpha}} \partial_{\gamma} \partial_{\acute{\gamma}}
+\eta_{\gamma\acute{\gamma}} \partial_{\alpha} \partial_{\acute{\alpha}})
\}
A^{a}_{\gamma\acute{\gamma}}
\eeqa
and is invariant with respect to the larger gauge transformations
$
\delta A^{a}_{\mu \lambda} =\partial_{\mu} \xi^{a}_{\lambda}+
\partial_{\lambda} \eta^{a}_{\mu},
$
where $\xi^{a}_{\lambda}$ and $\eta^{a}_{\mu}$ are eight arbitrary functions.
In the momentum representation the quadratic form is
\beqa\label{quadraticform}
H^{tot}_{\alpha\acute{\alpha}\gamma\acute{\gamma}}(k)=
(-\eta_{\alpha\gamma}\eta_{\acute{\alpha}\acute{\gamma}}
+{1 \over 2}\eta_{\alpha\acute{\gamma}}\eta_{\acute{\alpha}\gamma}
+{1 \over 2}\eta_{\alpha\acute{\alpha}}\eta_{\gamma\acute{\gamma}})k^2
+\eta_{\alpha\gamma}k_{\acute\alpha}k_{\acute{\gamma}}
+\eta_{\acute\alpha \acute{\gamma}}k_{\alpha}k_{\gamma}\nn\\
-{1 \over 2}(\eta_{\alpha\acute{\gamma}}k_{\acute\alpha}k_{\gamma}
+\eta_{\acute\alpha\gamma}k_{\alpha}k_{\acute{\gamma}}
+\eta_{\alpha\acute\alpha}k_{\gamma}k_{\acute{\gamma}}
+\eta_{\gamma\acute{\gamma}}k_{\alpha}k_{\acute\alpha}).
\eeqa

Let us consider also the symmetries of the remaining two terms in the
full Lagrangian ${{\cal L}}=  {{\cal L}}_1 +  {{\cal L}}_2 +  {{\cal L}}^{'}_2 $.
They have the form
$$
-{1\over 4}G^{a}_{\mu\nu}G^{a}_{\mu\nu,\lambda\lambda}
+{1\over 2} G^{a}_{\mu\nu}G^{a}_{\mu\lambda,\nu\lambda}.
$$
The part which is quadratic in fields  has the form
\beqa
{{\cal L}}^{free}_2 ={1 \over 2} A^{a}_{\alpha }\{&+&
(\eta_{\alpha\gamma} \partial^{2} - \partial_{\alpha} \partial_{\gamma} )
\eta_{\gamma^{'}\gamma^{''}} -
\eta_{\alpha\gamma}\partial_{\gamma^{'}} \partial_{\gamma^{''}}\nn\\
&-&{1 \over 2}
( \eta_{\gamma\gamma^{''}}\partial^{2}-\partial_{\gamma} \partial_{\gamma^{''}} )
\eta_{\alpha\gamma^{'}}
-{1 \over 2}
( \eta_{\gamma\gamma^{'}}\partial^{2}-\partial_{\gamma} \partial_{\gamma^{'}} )
\eta_{\alpha\gamma^{''}}\nn\\
&+&{1 \over 2}\eta_{\gamma\gamma^{''}}\partial_{\alpha} \partial_{\gamma^{'}}
+{1 \over 2}\eta_{\gamma\gamma^{'}}\partial_{\alpha} \partial_{\gamma^{''}}~
\}A^{a}_{\gamma\gamma^{'}\gamma^{''}} \nn\\&=&
{1 \over 2} A^{a}_{\alpha }
H_{\alpha \gamma\gamma^{'}\gamma^{''} } A^{a}_{\gamma\gamma^{'}\gamma^{''} } ,
\eeqa
where the quadratic form in the momentum representation is
\beqa
H_{\alpha \gamma\gamma^{'}\gamma^{''} }(k)=
&-&~~(\eta_{\alpha\gamma}k^2  -k_{\alpha}k_{\gamma})\eta_{\gamma^{'}\gamma^{''}}~
+~~\eta_{\alpha\gamma}k_{\gamma^{'}}k_{\gamma^{''}}\nn\\
&+&{1 \over 2}
( \eta_{\gamma\gamma^{''}}k^{2}-k_{\gamma} k_{\gamma^{''}} )
\eta_{\alpha\gamma^{'}}
-{1 \over 2}\eta_{\gamma\gamma^{''}}k_{\alpha} k_{\gamma^{'}}
\nn\\
&+&{1 \over 2}
( \eta_{\gamma\gamma^{'}}k^{2}-k_{\gamma} k_{\gamma^{'}} )
\eta_{\alpha\gamma^{''}}
-{1 \over 2}\eta_{\gamma\gamma^{'}}k_{\alpha} k_{\gamma^{''}}.
\eeqa
As one can see all divergences are equal to zero
\be\label{divergencestheird}
k_{\alpha}H_{\alpha \gamma\gamma^{'}\gamma^{''} }(k)=
k_{\gamma}H_{\alpha \gamma\gamma^{'}\gamma^{''} }(k)=
k_{\gamma^{'}}H_{\alpha \gamma\gamma^{'}\gamma^{''} }(k)=
k_{\gamma^{''}}H_{\alpha \gamma\gamma^{'}\gamma^{''} }(k)=0.
\ee
This result means that the quadratic part of the full Lagrangian
${{\cal L}}=  {{\cal L}}_1 +  {{\cal L}}_2 +  {{\cal L}}^{'}_2 $ is
invariant under the following local gauge transformations
\beqa\label{freedoublepolygaugesymmetric}
\tilde{\delta}_{\eta} A^{a}_{\mu} &=&  \partial_{\mu}\eta^a +... \nonumber\\
\tilde{\delta}_{\eta} A^{a}_{\mu\nu} &=&   \partial_{\nu}
\eta^{a}_{\mu} +...,\\
\tilde{\delta}_{\eta} A^{a}_{\mu\nu\lambda} &=& \partial_{\nu}
\eta^{a}_{\mu\lambda} + \partial_{\lambda}
\eta^{a}_{\mu\nu} +...\nn\\
........&.&......................................,\nn
\eeqa
in addition to the initial local gauge transformations (\ref{polygauge})
\beqa\label{firstlocal}
\delta_{\xi}  A^{a}_{\mu} &=& \partial_{\mu}\xi^{a}+...  \nonumber\\
\delta_{\xi}  A^{a}_{\mu\nu} &=& \partial_{\mu}\xi^{a}_{\nu}+...\nonumber\\
\delta_{\xi}  A^{a}_{\mu\nu\lambda} &=& \partial_{\mu}\xi^{a}_{\nu\lambda}+....
\eeqa

It is important to known how the transformation (\ref{freedoublepolygaugesymmetric}) looks like
when the gauge coupling constant is not equal to zero. The existence of the full
transformation is guaranteed by the conservation of the corresponding currents
(\ref{vectorcurrent}), (\ref{tensorcurrent}) and (\ref{tensorcurrentthierd}).
At the moment we can only guess
the full form of the second local gauge transformation requiring the closure of the
corresponding algebra. The extension we have found has the form \cite{Savvidy:2005fi}:
\beqa\label{doublepolygaugesymmetric}
\tilde{\delta}_{\eta} A^{a}_{\mu} &=& ( \delta^{ab}\partial_{\mu}
+g f^{acb}A^{c}_{\mu})\eta^b ,\\
\tilde{\delta}_{\eta} A^{a}_{\mu\nu} &=&  ( \delta^{ab}\partial_{\nu}
+  g f^{acb}A^{c}_{\nu})\eta^{b}_{\mu} + g f^{acb}A^{c}_{\mu\nu}\eta^{b},\nn\\
\tilde{\delta}_{\eta} A^{a}_{\mu\nu\lambda} &=& ( \delta^{ab}\partial_{\nu}
+g f^{acb} A^{c}_{\nu})\eta^{b}_{\mu\lambda} +( \delta^{ab}\partial_{\lambda}
+g f^{acb} A^{c}_{\lambda})\eta^{b}_{\mu\nu} %\nn\\&~&~~~~~~~~~~~~~~~~~~~~~~~~~~~+
+g f^{acb}(  A^{c}_{\mu  \nu}\eta^{b}_{\lambda }+
A^{c}_{\mu \lambda }\eta^{b}_{ \nu}+
A^{c}_{\mu\nu\lambda}\eta^{b})\nn\\
........&.&......................................,\nn
\eeqa
and forms a closed algebraic structure. The composition law of the gauge parameters
$\{ \eta,\eta_{\nu},\eta_{\nu\lambda},... \}$ is the same as in (\ref{gaugealgebra}).

\section{Enhanced Symmetry of Total Lagrangian and Equation of Motion}

In summary, we have the following Lagrangian for the
lower-rank tensor gauge fields:
\beqa\label{totalactiontwo}
{{\cal L}}=  {{\cal L}}_1 +  {{\cal L}}_2 +  {{\cal L}}^{'}_2 =
&-&{1\over 4}G^{a}_{\mu\nu}G^{a}_{\mu\nu}\\
&-&{1\over 4}G^{a}_{\mu\nu,\lambda}G^{a}_{\mu\nu,\lambda}
-{1\over 4}G^{a}_{\mu\nu}G^{a}_{\mu\nu,\lambda\lambda}\nn\\
&+&{1\over 4}G^{a}_{\mu\nu,\lambda}G^{a}_{\mu\lambda,\nu}
+{1\over 4}G^{a}_{\mu\nu,\nu}G^{a}_{\mu\lambda,\lambda}
+{1\over 2}G^{a}_{\mu\nu}G^{a}_{\mu\lambda,\nu\lambda}.\nn
\eeqa
Let us consider the equations of motion which follow from this Lagrangian for
the vector gauge field $A^{a}_{\nu}$:
\beqa\label{equationforfirstranktensor}
\nabla^{ab}_{\mu}G^{b}_{\mu\nu}
&+&{1\over 2 }\nabla^{ab}_{\mu} (G^{b}_{\mu\nu,\lambda\lambda}
+ G^{b}_{\nu\lambda,\mu\lambda}
+ G^{b}_{\lambda\mu,\nu\lambda})
+ g f^{acb} A^{c}_{\mu\lambda} G^{b}_{\mu\nu,\lambda}\\
&-&{1\over 2 }g f^{acb} (A^{c}_{\mu\lambda} G^{b}_{\mu\lambda,\nu}
+A^{c}_{\mu\lambda} G^{b}_{\lambda\nu,\mu}
+A^{c}_{\lambda\lambda} G^{b}_{\mu\nu,\mu}
+A^{c}_{\mu\nu} G^{b}_{\mu\lambda,\lambda})\nn\\
&+&{1\over 2 }g f^{acb} (
A^{c}_{\mu\lambda\lambda} G^{b}_{\mu\nu}
+ A^{c}_{\mu \mu\lambda } G^{b}_{\nu \lambda}
+ A^{c}_{\mu\nu\lambda} G^{b}_{\lambda\mu})\nn
=0
\eeqa\label{equationforsecondranktensor}
and for the second-rank tensor gauge field $A^{a}_{\nu\lambda}$:
\beqa\label{secondrankfieldequations}
&&\nabla^{ab}_{\mu}G^{b}_{\mu\nu,\lambda}
-{1\over 2} (\nabla^{ab}_{\mu}G^{b}_{\mu\lambda,\nu}
+\nabla^{ab}_{\mu}G^{b}_{\lambda\nu,\mu}
+\nabla^{ab}_{\lambda}G^{b}_{\mu\nu,\mu}
+\eta_{\nu\lambda} \nabla^{ab}_{\mu}G^{b}_{\mu\rho,\rho})\nn\\
&+&g f^{acb} A^{c}_{\mu\lambda} G^{b}_{\mu\nu} -
{1\over 2}g f^{acb}(A^{c}_{\mu\nu} G^{b}_{\mu\lambda}
+A^{c}_{\mu\mu} G^{b}_{\lambda\nu}
+A^{c}_{\lambda\mu} G^{b}_{\mu\nu}
+\eta_{\nu\lambda}  A^{c}_{\mu\rho} G^{b}_{\mu\rho})
=0.
\eeqa
The variation of the action with respect to the third-rank gauge field
$A^{a}_{\nu\lambda\rho}$ will give the equations
\be
\eta_{\lambda\rho}\nabla^{ab}_{\mu}G^{b}_{\mu\nu}-{1\over 2}
(\eta_{\nu\rho}\nabla^{ab}_{\mu}G^{b}_{\mu\lambda}  +
\eta_{\lambda\nu}\nabla^{ab}_{\mu}G^{b}_{\mu\rho}) +
{1\over 2} (\nabla^{ab}_{\rho}G^{b}_{\nu\lambda}  +
\nabla^{ab}_{\lambda}G^{b}_{\nu\rho})=0.
\ee
Representing these system of equations in the form
\beqa\label{perturbativeform}
\partial_{\mu} F^{a}_{\mu\nu}
+{1\over 2 }\partial_{\mu} (F^{a}_{\mu\nu,\lambda\lambda}
+ F^{a}_{\nu\lambda,\mu\lambda}
+ F^{a}_{\lambda\mu,\nu\lambda})= j^{a}_{\nu}\\
\partial_{\mu} F^{a}_{\mu\nu,\lambda}
-{1\over 2} (\partial_{\mu} F^{a}_{\mu\lambda,\nu}
+\partial_{\mu} F^{a}_{\lambda\nu,\mu}
+\partial_{\lambda}F^{a}_{\mu\nu,\mu}
+\eta_{\nu\lambda} \partial_{\mu}F^{a}_{\mu\rho,\rho}) = j^{a}_{\nu\lambda}\nn\\
\eta_{\lambda\rho}\partial_{\mu}F^{a}_{\mu\nu}-{1\over 2}
(\eta_{\nu\rho}\partial_{\mu}F^{a}_{\mu\lambda}  +
\eta_{\nu\lambda}\partial_{\mu}F^{a}_{\mu\rho}) +
{1\over 2} (\partial_{\rho}F^{a}_{\nu\lambda}  +
\partial_{\lambda}F^{a}_{\nu\rho})=
j^{a}_{\nu\lambda\rho},\nn
\eeqa
where $F^{a}_{\mu\nu} =  \partial_{\mu} A^{a}_{\nu  } -
\partial_{\nu} A^{a}_{\mu },~
F^{a}_{\mu\nu,\lambda} = \partial_{\mu} A^{a}_{\nu \lambda} -
\partial_{\nu} A^{a}_{\mu \lambda},~
F^{a}_{\mu\nu,\lambda\rho} = \partial_{\mu} A^{a}_{\nu \lambda\rho} -
\partial_{\nu} A^{a}_{\mu \lambda\rho}$ ,
we can find the corresponding conserved currents
\beqa\label{vectorcurrent}
j^{a}_{\nu } = &-&g f^{abc} A^{b}_{\mu } G^{c}_{\mu\nu }
-g f^{abc}\partial_{\mu} (A^{b}_{\mu } A^{c}_{\nu })\\
&-&{1\over 2 }g f^{abc}A^{b}_{\mu} (G^{c}_{\mu\nu,\lambda\lambda}
+ G^{c}_{\nu\lambda,\mu\lambda}
+ G^{c}_{\lambda\mu,\nu\lambda})
-{1\over 2 }\partial_{\mu} (I^{a}_{\mu\nu,\lambda\lambda}
+ I^{a}_{\nu\lambda,\mu\lambda}
+ I^{a}_{\lambda\mu,\nu\lambda})\nn\\
&-& g f^{abc} A^{b}_{\mu\lambda} G^{c}_{\mu\nu,\lambda}
+ {1\over 2 }g f^{abc} (A^{b}_{\mu\lambda} G^{c}_{\mu\lambda,\nu}
+A^{b}_{\mu\lambda} G^{c}_{\lambda\nu,\mu}
+A^{b}_{\lambda\lambda} G^{c}_{\mu\nu,\mu}
+A^{b}_{\mu\nu} G^{c}_{\mu\lambda,\lambda}) \nn\\
&-&{1\over 2} g f^{abc} (A^{b}_{\mu\lambda\lambda} G^{c}_{\mu\nu}
+A^{b}_{\lambda\mu\lambda} G^{c}_{\nu\mu}
+ A^{b}_{\mu\lambda\nu} G^{c}_{\lambda\mu}),\nn
\eeqa
where $I^{a}_{\mu\nu,\lambda\rho}=g f^{abc}(~A^{b}_{\mu}~A^{c}_{\nu\lambda\rho} +
 A^{b}_{\mu\lambda}~A^{c}_{\nu\rho}+A^{b}_{\mu\rho}~A^{c}_{\nu\lambda}
 + A^{b}_{\mu\lambda\rho}~A^{c}_{\nu} ~)$ and
\beqa\label{tensorcurrent}
j^{a}_{\nu\lambda}=&-&g f^{abc} A^{b}_{\mu} G^{c}_{\mu\nu,\lambda}
+{1\over 2 }g f^{abc} (A^{b}_{\mu} G^{c}_{\mu\lambda,\nu}
+A^{b}_{\mu} G^{c}_{\lambda\nu,\mu}
+A^{b}_{\lambda} G^{c}_{\mu\nu,\mu}
+\eta_{\nu\lambda}A^{b}_{\mu} G^{c}_{\mu\rho,\rho})\nn\\
&-&g f^{abc} A^{b}_{\mu\lambda} G^{c}_{\mu\nu} +
{1\over 2}g f^{abc}(A^{b}_{\mu\nu} G^{c}_{\mu\lambda}
+A^{b}_{\lambda\mu} G^{c}_{\mu\nu}
+A^{b}_{\mu\mu} G^{c}_{\lambda\nu}
+\eta_{\nu\lambda}  A^{b}_{\mu\rho} G^{c}_{\mu\rho})\nn\\
&-&g f^{abc} \partial_{\mu}
(A^{b}_{\mu} A^{c}_{\nu\lambda} + A^{b}_{\mu\lambda} A^{c}_{\nu}) +
{1\over 2}g f^{abc}
[\partial_{\mu}(A^{b}_{\mu} A^{c}_{\lambda\nu}+A^{b}_{\mu\nu} A^{c}_{\lambda})
+\partial_{\mu}(A^{b}_{\lambda} A^{c}_{\nu\mu}+A^{b}_{\lambda\mu} A^{c}_{\nu})\nn\\
&+&\partial_{\lambda} (A^{b}_{\mu} A^{b}_{\nu\mu}  + A^{b}_{\mu\mu} A^{c}_{\nu})
+\eta_{\nu\lambda}  \partial_{\mu}
(A^{b}_{\mu} A^{b}_{\rho\rho} + A^{b}_{\mu\rho} A^{c}_{\rho})],
\eeqa
\beqa\label{tensorcurrentthierd}
j^{a}_{\nu\lambda\rho}=&-&\eta_{\lambda\rho} ~g f^{abc} A^{b}_{\mu} G^{c}_{\mu\nu}
+{1\over 2 }g f^{abc} (\eta_{\nu\rho} A^{b}_{\mu} G^{c}_{\mu\lambda}
+\eta_{\nu\lambda}A^{b}_{\mu} G^{c}_{\mu\rho}
-A^{b}_{\rho} G^{c}_{\nu\lambda}
-A^{b}_{\lambda} G^{c}_{\nu\rho})\\
&-&\eta_{\lambda\rho} ~g f^{abc} \partial_{\mu}
(A^{b}_{\mu} A^{c}_{\nu}) +
{1\over 2}g f^{abc}
[\partial_{\mu}(\eta_{\nu\lambda} A^{b}_{\mu} A^{c}_{\rho}
+ \eta_{\nu\rho} A^{b}_{\mu} A^{c}_{\lambda})
-\partial_{\lambda} (A^{b}_{\nu} A^{c}_{\rho})
-\partial_{\rho}(A^{b}_{\nu} A^{c}_{\lambda})].\nn
\eeqa
Thus
\beqa
\partial_{\nu} j^{a}_{\nu}&=&0,~~~\nn\\
\partial_{\nu} j^{a}_{\nu\lambda}&=&0,~~~~
\partial_{\lambda} j^{a}_{\nu\lambda}=0,\nn\\
\partial_{\nu} j^{a}_{\nu\lambda\rho}&=&0,~~~~
\partial_{\lambda} j^{a}_{\nu\lambda\rho}=0,~~~~
\partial_{\rho} j^{a}_{\nu\lambda\rho}=0,
\eeqa
because, as we demonstrated above, the partial derivatives of the l.h.s. of the equations
(\ref{perturbativeform}) are equal to zero (see equations (\ref{zeroderivatives})
and equations (\ref{divergencestheird}) or calculate derivatives of l.h.s. of the
equations (\ref{perturbativeform}) ).

\section{Linearized Equations and Propagating Modes}

At the linearized level, when the gauge coupling constant g is equal to zero,
the equations of motion (\ref{secondrankfieldequations}) for the second-rank tensor
gauge fields will take the form
\beqa\label{mainequation}
\partial^{2}(A^{a}_{\nu\lambda} -{1\over 2}A^{a}_{\lambda\nu})
-\partial_{\nu} \partial_{\mu}  (A^{a}_{\mu\lambda}-
{1\over 2}A^{a}_{\lambda\mu} )&-&
\partial_{\lambda} \partial_{\mu}  (A^{a}_{\nu\mu} - {1\over 2}A^{a}_{\mu\nu} )
+\partial_{\nu} \partial_{\lambda} ( A^{a}_{\mu\mu}-{1\over 2}A^{a}_{\mu\mu})\nn\\
&+&{1\over 2}\eta_{\nu\lambda} ( \partial_{\mu} \partial_{\rho}A^{a}_{\mu\rho}
-  \partial^{2}A^{a}_{\mu\mu})=0
\eeqa
and, as we shall see below, they describe the propagation of massless particles
of spin 2 and spin 0. It is also easy to see that for the symmetric
part of the tensor gauge field
$(A^{a}_{\nu\lambda} + A^{a}_{\lambda\nu})/2$ our equation
reduces to the well known Fierz-Pauli-Schwinger-Chang-Singh-Hagen-Fronsdal equation
\be\label{fierz}
\partial^{2} A_{\nu\lambda}
-\partial_{\nu} \partial_{\mu}  A_{\mu\lambda} -
\partial_{\lambda} \partial_{\mu}  A_{\mu\nu}
+ \partial_{\nu} \partial_{\lambda}  A_{\mu\mu}
+\eta_{\nu\lambda}  (\partial_{\mu} \partial_{\rho}A_{\mu\rho}
- \partial^{2} A_{\mu\mu}) =0,
\ee
which describes the propagation of massless tensor boson with two
physical polarizations, the $\lambda= \pm 2$ helicity states. For the
antisymmetric part $(A^{a}_{\nu\lambda} - A^{a}_{\lambda\nu})/2$
the equation reduces to the form
\be\label{antisymmetric}
\partial^{2} A_{\nu\lambda}
-\partial_{\nu} \partial_{\mu}  A_{\mu\lambda} +
\partial_{\lambda} \partial_{\mu}  A_{\mu\nu}=0
\ee
and describes the propagation of massless scalar boson with
one physical polarization, the $\lambda= 0$ helicity state.

We can find out now how many propagating degrees of freedom
describe the system of equations (\ref{mainequation}) in the classical theory.
Taking the trace of the equation (\ref{mainequation}) we shall get
\be\label{freeequationtrace}
\partial_{\mu} \partial_{\rho}A^{a}_{\mu\rho}
-  \partial^{2}A^{a}_{\rho\rho} =0,
\ee
and the equation (\ref{mainequation}) takes the form
\beqa\label{tracelessequations}
\partial^{2}(A^{a}_{\nu\lambda} -{1\over 2}A^{a}_{\lambda\nu})
-\partial_{\nu} \partial_{\mu}  (A^{a}_{\mu\lambda}-
{1\over 2}A^{a}_{\lambda\mu} )-
\partial_{\lambda} \partial_{\mu}  (A^{a}_{\nu\mu} - {1\over 2}A^{a}_{\mu\nu} )
+{1\over 2}\partial_{\nu} \partial_{\lambda} A^{a}_{\mu\mu}
=0 .
\eeqa
Using the gauge invariance (\ref{largegaugetransformation}) we can impose the
Lorentz invariant supplementary
conditions on the second-rank gauge fields $A_{\mu\lambda}$:
$
\partial_{\mu} A^{a}_{\mu\lambda} =a_{\lambda} ,~~
\partial_{\lambda} A^{a}_{\mu\lambda} =b_{\mu} ,
$
where $a_{\lambda}$ and $b_{\mu}$ are arbitrary functions, or
one can suggest alternative
supplementary conditions in which the quadratic form
(\ref{totalfreelagrangian}), (\ref{totfreelagrangianalternativeform}),
(\ref{quadraticform}) is diagonal:
\be\label{diagonalgauge}
\partial_{\mu} A^{a}_{\mu\lambda} -{1\over 2} \partial_{\lambda} A^{a}_{\mu\mu}=0,~~
\partial_{\lambda} A^{a}_{\mu\lambda} -{1\over 2} \partial_{\mu} A^{a}_{\lambda\lambda}=0.
\ee
In this gauge the equation (\ref{tracelessequations}) has the  form
\beqa\label{gaugefixedequations}
\partial^{2} A^{a}_{\nu\lambda}  =0
\eeqa
and in the momentum representation
$
A_{\mu\nu}(x) = e_{\mu\nu}(k) e^{ikx}
$
from  equation (\ref{gaugefixedequations}) it follows that $k^2=0$ and
we have {\it massless particles}.

For the symmetric part of the tensor field $A^{a}_{\mu\lambda}$ the supplementary
conditions (\ref{diagonalgauge}) are equivalent to the harmonic gauge
\be\label{harmonic}
\partial_{\mu} (A^{a}_{\mu\lambda} + A^{a}_{\lambda\mu})
 -{1\over 2} \partial_{\lambda}( A^{a}_{\mu\mu}+A^{a}_{\mu\mu})=0,
\ee
and the residual gauge transformations are defined by the gauge parameters
$ \xi^{a}_{\lambda}+\eta^{a}_{\lambda}$ which should satisfy the equation
\beqa\label{residual1}
\partial^{2}(\xi^{a}_{\lambda}+\eta^{a}_{\lambda})=0.
\eeqa
Thus imposing the harmonic gauge (\ref{harmonic}) and
using the residual gauge transformations
(\ref{residual1}) one can see that the number of propagating physical polarizations
which are described by
the symmetric part of the tensor field $A^{a}_{\mu\lambda}$ are given by two
helicity states $\lambda= \pm 2$ multiplied by the dimension of the group G (a=1,...,N).

For the anisymmetric part of the tensor field $A^{a}_{\mu\lambda}$ the supplementary
conditions (\ref{diagonalgauge}) are equivalent to the Lorentz gauge
\be
\partial_{\mu}(A^{a}_{\mu\lambda} - A^{a}_{\lambda\mu})=0
\ee
and together with the equation of motion they describe the propagation of one
physical polarization of helicity $\lambda= 0$ multiplied by the dimension of
the group G (a=1,...,N).

Thus we have seen that the extended gauge symmetry (\ref{largegaugetransformation})
with eight gauge parameters is sufficient to
exclude all negative norm polarizations from the spectrum of the second-rank
{\it nonsymmetric tensor gauge field} $A_{\mu\lambda}$ which describes now
the propagation of three physical modes of helicities $\pm 2$ and $0$.

In the  gauge (\ref{diagonalgauge})  we shall get
$$
H^{fix}_{\alpha\acute{\alpha}\gamma\acute{\gamma}}(k) =
(\eta_{\alpha\gamma}\eta_{\acute{\alpha}\acute{\gamma}}
-{1\over 2}\eta_{\alpha\acute{\gamma}}\eta_{\acute{\alpha}\gamma}
-{1\over 4}\eta_{\alpha\acute{\alpha}}\eta_{\gamma\acute{\gamma}})(-k^2 )
$$
and the propagator $\Delta_{\gamma\acute{\gamma}\lambda\acute{\lambda}}(k)$
from the equation
$
H^{fix}_{\alpha\acute{\alpha}\gamma\acute{\gamma}}(k)
\Delta_{\gamma\acute{\gamma}\lambda\acute{\lambda}}(k) =
\eta_{\alpha\lambda}\eta_{\acute{\alpha}\acute{\lambda}}~~,
$
thus
\be
\Delta_{\gamma\acute{\gamma}\lambda\acute{\lambda}}(k) = -
{4 \eta_{\gamma\lambda}\eta_{\acute{\gamma}\acute{\lambda}}
+2\eta_{\gamma \acute{\lambda} } \eta_{ \acute{\gamma} \lambda}
-3\eta_{\gamma\acute{\gamma}}\eta_{\lambda\acute{\lambda}}
 \over 3(k^2 - i\varepsilon)}~~.
\ee
The corresponding residue can be represented as a sum
\beqa
{4 \eta_{\gamma\lambda}\eta_{\acute{\gamma}\acute{\lambda}}
+2\eta_{\gamma \acute{\lambda} } \eta_{ \acute{\gamma} \lambda}
-3\eta_{\gamma\acute{\gamma}}\eta_{\lambda\acute{\lambda}}
 \over 3 }
=&+& (\eta_{\gamma\lambda}\eta_{\acute{\gamma}\acute{\lambda}}
+\eta_{\gamma\acute{\lambda}}\eta_{\acute{\gamma}\lambda}
-\eta_{\gamma\acute{\gamma}}\eta_{\lambda\acute{\lambda}})+
 {1\over 3}(\eta_{\gamma\lambda}\eta_{\acute{\gamma}\acute{\lambda}}
-\eta_{\gamma\acute{\lambda}}\eta_{\acute{\gamma}\lambda}).\nn
\eeqa
The first term describes the $\lambda= \pm 2$ helicity states and is
represented by the symmetric part of the polarization tensor $e_{\mu\lambda}$,
the second term describes $\lambda= 0$ helicity state and is represented
by its antisymmetric part.
Indeed, for the massless case, when $k_{\mu}$ is aligned along the third axis,
$k_{\mu}= (k,0,0,k)$, we have two independent polarizations of the helicity-2
particle and spin-zero {\it axion}
\beqa
e^{1}_{\mu\lambda}={1\over \sqrt{2}}
\left( \begin{array}{llll}
  0,0,~~0,0\\
  0,1,~~0,0\\
  0,0,-1,0\\
  0,0,~~0,0
\end{array} \right), e^{2}_{\mu\lambda}={1\over \sqrt{2}}
\left( \begin{array}{ll}
  0,0,0,0\\
  0,0,1,0\\
  0,1,0,0\\
  0,0,0,0
\end{array} \right),
e^{A}_{\mu\lambda}={1\over \sqrt{2}}
\left( \begin{array}{ll}
  0,~~0,0,0\\
  0,~~0,1,0\\
  0,-1,0,0\\
  0,~~0,0,0
\end{array} \right),\nn\\
\eeqa
with the property that
$
e^{1}_{\gamma\acute{\gamma}}e^{1}_{\lambda\acute{\lambda}}  +
e^{2}_{\alpha\acute{\alpha}} e^{2}_{\gamma\acute{\gamma}}\simeq
{1\over 2}(\eta_{\gamma\lambda}\eta_{\acute{\gamma}\acute{\lambda}}
+\eta_{\gamma\acute{\lambda}}\eta_{\acute{\gamma}\lambda}
-\eta_{\gamma\acute{\gamma}}\eta_{\lambda\acute{\lambda}})
$
and
$
e^{A}_{\gamma\acute{\gamma}}e^{A}_{\lambda\acute{\lambda}}
\simeq{1\over 2}(\eta_{\gamma\lambda}\eta_{\acute{\gamma}\acute{\lambda}}
-\eta_{\gamma\acute{\lambda}}\eta_{\acute{\gamma}\lambda}).
$
The symbol $\simeq$ means that the equation holds up to longitudinal terms.

Thus the general second-rank tensor gauge field with 16
components $A_{\mu\lambda}$ describes in this theory three
physical propagating massless polarizations.

\section{Higher-Spin Extension of Electroweak Theory}

Let us consider the possible extension of the standard model of electroweak
interactions which follows from the above generalization. In the first model
which we shall consider only the $SU(2)_L$ group will be extended to higher spins,
but not the $U(1)_Y$ group.  The  $W^{\pm},Z$  gauge bosons will
receive their higher-spin descendence
$$
(W^{\pm},Z)_{\mu},~~~~~~~~~(\tilde{W}^{\pm},\tilde{Z})_{\mu\lambda}, .....
$$
and the doublet of complex Higgs
scalars will appear together with their higher-spin partners:
$$
(\begin{array}{c}
  \phi^+ \\
   \phi^o
\end{array})~~, ~~~~~~~(\begin{array}{c}
  \phi^+ \\
   \phi^o
\end{array})_{\lambda}~~, ~~~~~~~(\begin{array}{c}
  \phi^+ \\
   \phi^o
\end{array})_{\lambda\rho}~~,......~~~~~~~~~~~~~~~Y=+1.
$$
The Lagrangian which describes  the interaction of the tensor gauge bosons with
scalar fields and tensor bosons is:
\beqa\label{standardmodellagrangian}
{{\cal L}}  =&-&{1\over 4}G^{i}_{\mu\nu}
G^{i}_{\mu\nu} - {1\over 4}F_{\mu\nu}
F_{\mu\nu} + (\partial_{\mu}   + {ig^{'} \over 2} B_{\mu} +
{ig \over 2}\tau^{i} A^{i}_{\mu} )\phi^{\dag} ~
(\partial_{\mu}   - {ig^{'} \over 2}B_{\mu} - {ig \over 2}\tau^{i}A^{i}_{\mu} )\phi \nn\\
&-&{1\over 4}G^{i}_{\mu\nu,\lambda}G^{i}_{\mu\nu,\lambda}
-{1\over 4}G^{i}_{\mu\nu}G^{i}_{\mu\nu,\lambda\lambda}
+{1\over 4}G^{i}_{\mu\nu,\lambda}G^{i}_{\mu\lambda,\nu}
+{1\over 4}G^{i}_{\mu\nu,\nu}G^{i}_{\mu\lambda,\lambda}
+{1\over 2}G^{i}_{\mu\nu}G^{i}_{\mu\lambda,\nu\lambda}\nn\\
&+&{g^{2} \over 4} \phi^{\dag} \tau^{i} A^{i}_{\mu\lambda} \tau^{j} A^{j}_{\mu\lambda}\phi
+\nabla_{\mu}\phi^{\dag}_{\lambda} ~\nabla_{\mu}\phi_{\lambda} +
{1\over 2}\nabla_{\mu}\phi^{\dag}_{\lambda\lambda} ~ \nabla_{\mu}\phi +
{1\over 2}\nabla_{\mu}\phi^{\dag} ~ \nabla_{\mu}\phi_{\lambda\lambda} \nonumber\\
&-& ig \nabla_{\mu}\phi^{\dag} ~A_{\mu\lambda}\phi_{\lambda} +
ig \phi^{\dag}_{\lambda} A_{\mu\lambda}~\nabla_{\mu}\phi -
ig \nabla_{\mu}\phi^{\dag}_{\lambda} ~A_{\mu\lambda}\phi+
ig \phi^{\dag} A_{\mu\lambda}  ~ \nabla_{\mu}\phi_{\lambda}-\nn\\
&-&{1\over 2}ig \nabla_{\mu}\phi^{\dag} ~A_{\mu\lambda\lambda}\phi +
{1\over 2}ig \phi^{\dag} A_{\mu\lambda\lambda}~\nabla_{\mu}\phi ~,
\eeqa
where
$
\nabla_{\mu}  = \partial_{\mu}   - {ig^{'} \over 2} Y B_{\mu} - ig  T^{i} A^{i}_{\mu},
$
Y  is hypercharge, Q is charge, $Q = T_3 + Y/2$, and for isospinor fields
$T^i = \tau^i /2 $. The three terms in the first line represent the
standard electroweak model and the rest
of the terms - its higher-spin generalization. Therefore all parameters of the standard
model are incorporated in the extension.
The first  term in the third line will
generate the masses of the tensor $\tilde{W}^{\pm},\tilde{Z}$  gauge bosons:
\be
{1\over 8} g^2 \eta^2 [(A^{3}_{\mu\lambda})^2 +
2 A^{+}_{\mu\lambda}A^{-}_{\mu\lambda}],
\ee
when the scalar fields acquire  the vacuum expectation value $\eta$:
$
\phi = {1\over \sqrt{2}} (\begin{array}{c}
                           0 \\
                           \eta + \chi(x)
                         \end{array})
$
and
$$
\tilde{Z}_{\mu\lambda} =A^{3}_{\mu\lambda},~~~   \tilde{W}^{\pm}_{\mu\lambda}=
{1\over \sqrt{2}}(A^{1}_{\mu\lambda} \pm  i A^{2}_{\mu\lambda}),
$$
Thus all three intermediate spin-2 bosons will acquire the same mass
$
m_{\tilde{W},\tilde{Z}} = {1\over 2} g \eta = m_W~~.
$
The rest of the
terms describe the interaction between "old" and new particles.
One should also introduce
the Yukawa self-interaction  for the bosons $\phi_{\lambda}$
in order to make them massive.

Let us consider the fermion sector of the extended electroweak model.
One should note that the interaction of tensor gauge bosons with fermions is not as
usual as one could expect. Indeed, let us now analyze the interaction with new
spinor-tensor leptons
$$
L = {1\over 2}(1+ \gamma_{5})(\begin{array}{c}
  \nu_e \\
   e
\end{array}),~~~~L_{\lambda} =
{1\over 2}(1+ \gamma_{5})(\begin{array}{c}
  \nu_e \\
   e
\end{array})_{\lambda},~~~~L_{\lambda\rho} =
{1\over 2}(1+ \gamma_{5})(\begin{array}{c}
  \nu_e \\
   e
\end{array})_{\lambda\rho}~..~~~Y=-1.
$$
All these left-handed states have hypercharge $Y=-1$ and the only right-handed
state
$$
R = {1\over 2}(1 - \gamma_{5})e, ~~~~~~~~~~~~~~~~~~~~Y=-2
$$
has the hypercharge $Y = -2$. The corresponding Lagrangian will take the form
\beqa
{{\cal L}}_{F}= \bar{L} \not\!\nabla L + \bar{R} \not\!\nabla R  +
 \bar{L}_\lambda \not\!\nabla L_\lambda + {1\over 2}\bar{L} \not\!\nabla L_{\lambda\lambda} +
{1\over 2}\bar{L}_{\lambda\lambda} \not\!\nabla L +
 g \bar{L}_\lambda \not\!\!A_{\lambda}  L  + g \bar{L} \not\!\!A_{\lambda}  L_\lambda
+ {1\over 2}g \bar{L} \not\!\!A_{\lambda\lambda}  L ,\nn
\eeqa
where the first two terms describe the standard electroweak interaction of
vector gauge bosons with standard spin-1/2 leptons,
the next three terms describe the interaction of the vector gauge bosons
with new leptons of the spin 3/2 and finally the last three terms describe
the interaction of the new tensor gauge bosons
$\tilde{W}^{\pm},\tilde{Z}$ with standard spin-1/2 and
spin-3/2 leptons.

The new interaction vertices generate
decay of the standard vector gauge bosons through the channels
$$
\gamma,Z \rightarrow e_{3/2}+\bar{e}_{3/2},~~~
\gamma,Z \rightarrow \nu_{3/2}+\bar{\nu}_{3/2},~~~
W \rightarrow \nu_{3/2} + e_{3/2},~~~W \rightarrow ~\nu_{3/2}+e_{3/2}~,
$$
where a pair of new leptons is created.
The observability of these channels depends on the masses of the new
leptons. This information is encoded into the Yukawa couplings, as it takes place
for the standard leptons of the spin 1/2. We can only say that they are large
enough not to be seen at low energies, but are predicted to be visible
at higher-energy experiments.

The decay reactions of the new tensor gauge bosons $\tilde{W}^{\pm},\tilde{Z}$
can take place through the channels
\be\label{decaychannels}
\tilde{Z}\rightarrow e_{3/2}+\bar{e}_{1/2},~~~
\tilde{Z}\rightarrow \nu_{3/2}+\bar{\nu}_{1/2},~~~
\tilde{W} \rightarrow ~\nu_{1/2}+e_{3/2},~~~
\tilde{W} \rightarrow ~\nu_{3/2}+e_{1/2}.
\ee
The main feature of these processes is that they create a pair which
consists of a standard lepton $e_{1/2}$ and of a new lepton $e_{3/2}$
of the spin 3/2. Because in all these
reactions there always participates a new lepton, they may
take place also at large enough energies, but it is impossible
to predict the threshold energy
because we do not know the corresponding Yukawa couplings. The
situation with Yukawa couplings is the same as it is in the standard model.
There is no decay channels of the new tensor bosons only into the standard leptons,
as one can see from the Lagrangian. Therefore it is also impossible to create
tensor gauge bosons directly in $e^+  + e^-$ annihilation, but they can appear in the
decay of the  $Z $
\be\label{mostpromising}
e^+  + e^- \rightarrow Z \rightarrow \tilde{W}^+ + \tilde{W}^- ~~
\ee
and will afterwards decay through the channels discussed
above (\ref{decaychannels}) $\tilde{W} \rightarrow ~\nu + \tilde{e}$ or
$\tilde{W} \rightarrow ~\tilde{\nu} + e$ .
{\it It seems that reaction (\ref{mostpromising}), predicted by the generalized
theory, is the most appropriate candidate which could be tested in
the experiment}. The details will be given in the forthcoming publication.

Let us consider now the tensor extension of the $U(1)_Y$, in that case we
shall have the massless spin-2 descendent of the photon,   which
we shall associated with the graviton. The right-handed
sector should be enlarged in the following way:
$$
R = {1\over 2}(1 - \gamma_{5})e, ~~~~R_{\lambda} = {1\over 2}(1 - \gamma_{5})e_{\lambda}, ~~~~
R_{\lambda\rho} = {1\over 2}(1 - \gamma_{5})e_{\lambda\rho},~~...,~~~~~~~~~~~~~~~Y=-2,
$$
and the Lagrangian will take the form
\beqa
{{\cal L}}_{F}&=& \bar{L} \not\!\nabla L + \bar{R} \not\!\nabla R \\
&+&\bar{L}_\lambda \not\!\nabla L_\lambda + {1\over 2}\bar{L} \not\!\nabla L_{\lambda\lambda} +
{1\over 2}\bar{L}_{\lambda\lambda} \not\!\nabla L +
g \bar{L}_\lambda \not\!\!A_{\lambda}  L  + g \bar{L} \not\!\!A_{\lambda}  L_\lambda
+ {1\over 2}g \bar{L} \not\!\!A_{\lambda\lambda}  L     \nn\\
&+&\bar{R}_\lambda \not\!\nabla R_\lambda + {1\over 2}\bar{R} \not\!\nabla R_{\lambda\lambda} +
{1\over 2}\bar{R}_{\lambda\lambda} \not\!\nabla R +
g^{'} \bar{R}_\lambda \not\!\!B_{\lambda}  R  + g^{'} \bar{R} \not\!\!B_{\lambda}  R_\lambda
+ {1\over 2}g^{'} \bar{R} \not\!\!B_{\lambda\lambda}  R,\nn
\eeqa
where the terms in the last line describe the interaction of the Abelian $U(1)_Y$
tensor fields $B_{\mu},B_{\mu\lambda},... $
with the right-handed sector of new leptons.

I wish to thank the organizers of the conference Corfu2005 and
especially Ionnis Bakas for the invitation.
This work was partially supported by the EEC Grant no. MRTN-CT-2004-005616.

\end{document}